# Colossal Dielectric Response and Electric Polarization in Lithium Nitrate

Na Du,[†] Yan Zhao,[†] Enting Xu,[†] Jianwei Han, Peng Ren,* Fei Yen*

*School of Science, Harbin Institute of Technology, Shenzhen, University Town, Shenzhen, Guangdong 518055, P. R. China.*

**Corresponding Author:**
*E-mail: fyen at hit.edu.cn, renpeng at hit.edu.cn
[†]These authors contributed equally to this work

**Abstract:** Lithium nitrate $LiNO_3$ is identified to possess a dielectric constant ε' larger than $6\times10^6$ at 1 kHz in powder samples above the critical temperature $T_W$ = 306 K. For single crystalline samples, ε' can be sustained to remain above $10^5$ and the dissipation factor below 10 in the temperature region of 280-340 K after a simple 'activation' process. Moreover, pyroelectric current measurements show $LiNO_3$ to be ferroelectric with an electric polarization of $P = \pm1,200$ μC/cm². Both ε' and $P$ are amongst one of the highest in all known materials. We propose a model suggesting the mechanism underlying the colossal magnitudes of ε' and $P$ to stem from a gearing-ungearing process of the planar $NO_3^-$ at the macroscopic level. Our results potentially push the boundaries of ceramic capacitors.

1. **Introduction**

Lithium nitrate LiNO$_3$ crystallizes in the trigonal structure [1-3]. According to Strømme [1] the packing of the rhombohedral lattice along with all the O–O distances requires that the planar NO$_3^-$ ions are spatially ordered. As such, it was suggested that LiNO$_3$ is the only univalent nitrate XNO$_3$ (where X = Li, Na, NH$_4$, K, Rb, Cs, Ag and Tl) that does not possess a spatially disordered solid phase so an order-disorder phase transition is not expected to occur. However, we recently observed a reversible solid-solid phase transition in LiNO$_3$ according to experimental measurements of the magnetic susceptibility [4]. It was believed that the librations (the in-plane, back-and-forth rocking) of neighboring NO$_3^-$ become geared below $T_C$ = 265 K and ungeared above $T_W$ = 303 K. Note the associated hysteresis spans nearly 40 K. Fermor & Kjekshus also reported on a phase transition at 263 K during cooling but their warming data stopped at 296 K, just short of $T_W$ [5]. The dielectric constant data of Fermor & Kjekshus were rather sparse, varying from 3 K to 10 K per data point and the sweeping speed was also not reported. It was therefore of general interest to re-investigate the dielectric properties of LiNO$_3$ in the range of 200–330 K under slow and constant sweeping rates.

To our surprise, the measured dielectric constant with respect to temperature $\varepsilon'(T)$ was observed to increase by around four orders of magnitude at $T_W$ when powder samples (pressed into disc pellets) were directly warmed from room temperature. This result renders LiNO$_3$ as having one of the highest (if not the highest at 1 kHz) dielectric constants ever reported. Materials with a giant dielectric constant (GDC) have potential to be employed as dielectrics for energy storage and actuators having nonlinear opto-electronic effects [6]. Many types of materials with GDC exist such as ceramics [7], high entropy alloys [8], plastic crystals [9], photo-induced organic-inorganic perovskites [10] and composite materials [11,12]. The mechanisms underlying GDC are also diverse: they include ferroelectricity [13,14], polaron hopping [15], grain boundary, electrode interfacial, and lattice-defect-electron-pinning effects [16-18].

In this work, we present data on ε'($T$) along with the imaginary part of the dielectric constant ε"($T$) of LiNO$_3$ in powder and single crystalline forms in the ranges of 25 Hz to 10 kHz and 200 to 340 K. The complex dielectric constant at 310 K in the frequency range of 20 Hz to 300 kHz is also presented. Large dielectric constants often correlate to ferroelectricity [19] so measurements of the pyroelectric current were also carried out. We find a record high electric polarization that can be switched according to the polarity of a *dc* external electric field characterizing LiNO$_3$ as ferroelectric. Lastly, we discuss a possible mechanism leading to the observed colossal dielectric response and electric polarization as stemming from an isostructural first-order phase transition involving the gearing-ungearing process of the NO$_3^-$ ions.

## 2. Materials and methods

Lithium nitrate (CAS# 7790-69-4) 99.99% in purity was purchased from Aladdin, Inc., Macklin, Inc. and Meryer, Inc. For the samples labelled 'powdered', the reagent was used 'as is' (with no further purification), taken directly from the bottle and pressed into disc pellets having diameters of 3.00 mm and thicknesses $d$ of 0.38 to 0.44 mm. Silver paint electrodes were applied onto the two flat surfaces of the pellets to form a parallel plate capacitor having platinum wire leads. The real ε' and imaginary ε" parts of the dielectric constant were obtained from the measured capacitance $C$ and dissipation factor tan δ, respectively, with an Agilent A4980E impedance analyzer. The following relationships were used to obtain the complex dielectric constant: $C = 0.0015^2 \pi$ ε' ε$_0$ / $d$ and tan δ = ε" / ε' where ε$_0$ is the permittivity of free space. Temperature control and thermometry were realized with the cryostat of a PPMS system manufactured by Quantum Design, Inc.

For the single crystalline samples, a solution of LiNO$_3$+H$_2$O was slowly evaporated. Large, flat, transparent and trigonal-shaped crystals were then harvested. The pyroelectric current was measured by a Keithley 4917B electrometer.

## 3. Results and discussion

### 3.1. Complex dielectric constant as a function of temperature

#### 3.1.1. LiNO₃ in powder form

Figures 1a and 1b show $\varepsilon'(T)$ and $\varepsilon''(T)$, respectively, of LiNO$_3$ in powder form under 1 kHz during warming from 300 K to 310 K from three providers. Near 306 K $\varepsilon'(T)$ increased dramatically by nearly four orders of magnitude in the span of 5 K to reach values near $7\times10^6$. The behavior of $\varepsilon''(T)$ followed the same pattern and possessed nearly the same order of magnitude at temperatures above 307 K. Figure 1c shows the same $\varepsilon'(T)$ scans but under different frequencies. $\varepsilon''(T)$ exhibited a similar profile so instead the dissipation factor (tan $\delta = \varepsilon'' / \varepsilon'$) is shown in Fig. 1d. The largest observed $\varepsilon'$ was ~$2\times10^8$ at 306.7 K under 25 Hz. The values in parenthesis in Fig. 1a show the thicknesses of the disc-shaped samples (inset of Fig. 1b). Given that $\varepsilon'$ was relatively independent of sample thickness indicates the colossal response of $\varepsilon'$ is mostly a bulk effect.

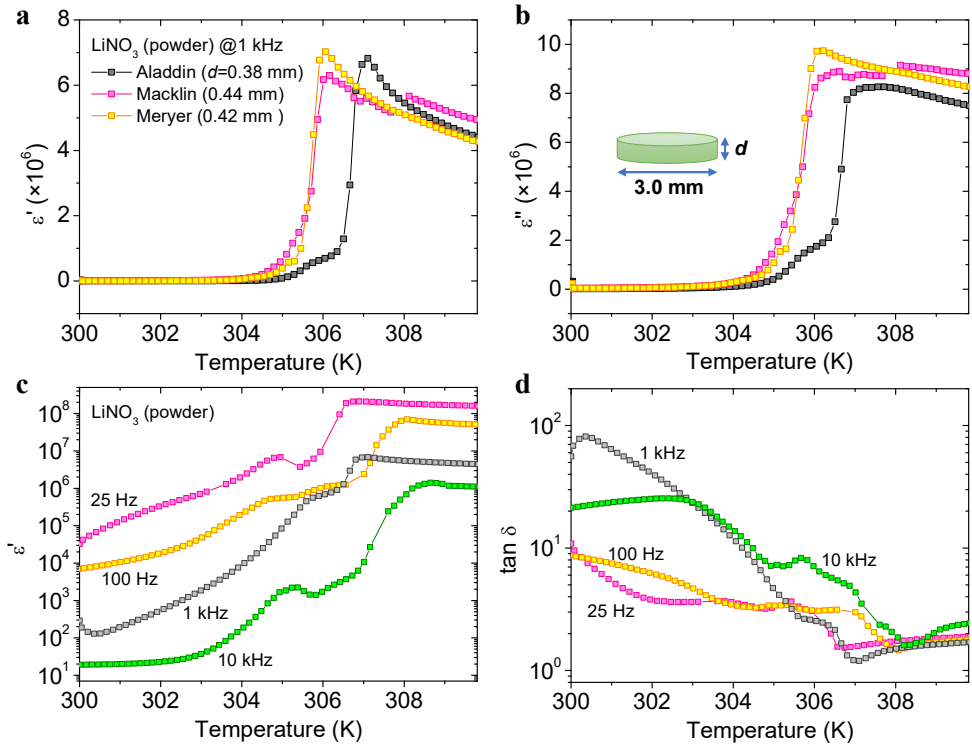

**Fig. 1.** (a) Real $\varepsilon'(T)$ and (b) Imaginary $\varepsilon''(T)$ parts of the dielectric constant of LiNO$_3$ (powder pressed into pellet) under 1 kHz during warming (1 K/min) from three manufacturers. Inset shows the dimensions and shape of the pellets. (c) $\varepsilon'(T)$ under different frequencies. (d) Dissipation factor tan $\delta = \varepsilon'' / \varepsilon'$ against temperature.

Figures 2a and 2b show $\varepsilon'(T)$ and $\varepsilon''(T)$ of one sample subjected to successive warming and cooling cycles under 1 kHz in a larger temperature range. In general, both $\varepsilon''(T)$ and $\varepsilon''(T)$ followed a similar behavior. Here, the most pronounced feature is again the sharp increase by orders of magnitude near $T_W$ during the warming runs. The sudden increase of $\varepsilon'(T)$ and $\varepsilon''(T)$ of the latter two warming runs were less pronounced but they occurred at nearly the same temperature. From the cooling curves in Figs. 2a and 2b, the disorder-order transition temperature can also be discerned to occur near $T_C = 275$ K where $\varepsilon'(T)$ and $\varepsilon''(T)$ exhibited a sudden drop by many orders of magnitude. After many repetitions with different samples, we found that $T_W$ varies only between 303.7 and 307.0 K while $T_C$ can vary between 262.0 to 276.5 K.

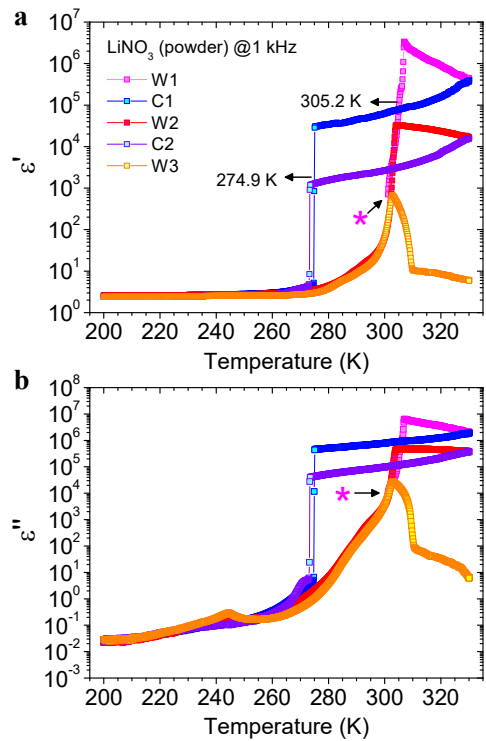

**Fig. 2.** (a) $\varepsilon'(T)$ and (b) $\varepsilon''(T)$ of LiNO$_3$ (powder) under 1 kHz subjected to three warming (W1, W2 and W3 and two cooling cycles (C1 and C2) all at 1 K/min. The * symbol indicates the start of the experiments.

### 3.1.2. LiNO$_3$ in crystalline form

Figures 3a and 3b show $\varepsilon'(T)$ and $\varepsilon''(T)$ under 1 kHz of a single crystal sample of LiNO$_3$ in the same temperature range as that of the powder samples. The magnitude of the dielectric response upon crossing $T_W$ was nearly the same. However, there were several key differences: 1) starting from 300 K (indicated by the asterisks in Figs. 3a and 3b) the sample seemed to start off in the disordered phase so no transition was observed during the first warming run W1. 2) During cooling, ε'($T$) and ε"($T$) increased first before exhibiting the same sharp drop off at $T_C$ (curves C1 and C2). 3) When warming back (curves W2 and W3) the magnitude of the dielectric response was always sustained to remain above $10^5$ upon crossing $T_W$; *i.e.* the magnitude of ε'($T$) did not deteriorate after each cycle. 4) If the temperature of the system remained above $T_C$, then the magnitude of ε'($T$) would also remain above $10^5$ (curves C3, W4 and C4).

It therefore seemed to us that for practical applications, once the temperature is warmed above $T_W$ to 'activate' the colossal magnitude of ε', its value can be kept to remain above $10^5$ as long as the temperature did not drop below $T_C$. Figures 3c and 3d show ε'($T$) and tan δ of the same crystal under different frequencies in the range between 280 K and 340 K. The cooling and warming rates were set faster to 3 K/min. Throughout this region, ε'($T$) remained stable and the values of tan δ, under all frequencies, were less than 10.

Sharp changes in ε'($T$) and ε"($T$) usually pertain to structural phase transitions. To rule out whether this is the case in $LiNO_3$, XRD measurements were performed on single crystalline samples at 100 K, 240 K and 380 K (results are shown in the Supplementary Material). At all three temperatures the crystal structure of $LiNO_3$ was confirmed to be trigonal (Space group No. 167) in agreement with existing literature [1-3]. From such, the large magnitudes of ε'($T$) and ε"($T$) appear to mainly represent the structural dynamics of the system; more specifically, the frequencies of the $NO_3^-$ librations at different temperatures. This will be discussed in more detail below. The packing of the unit lattice is also consistent with Strømme's conclusion that no *spatial* order-disorder takes place in the solid state of $LiNO_3$ [1]. Therefore, the most plausible scenario we can think of is to presume the observed phase transition to be associated to a *temporal* order-disorder (gearing-ungearing) of the $NO_3^-$ according to our previous results [4,20], *i.e.* the anions librate independently from each other in the disordered phase, and in contrast, the intermolecular librations become coherent in the ordered phase. In the Supplementary Material file, results of the measured specific heat across $T_W$ is also included. A small λ-peak anomaly was identified which further verifies the order-disorder nature of the phase transition.

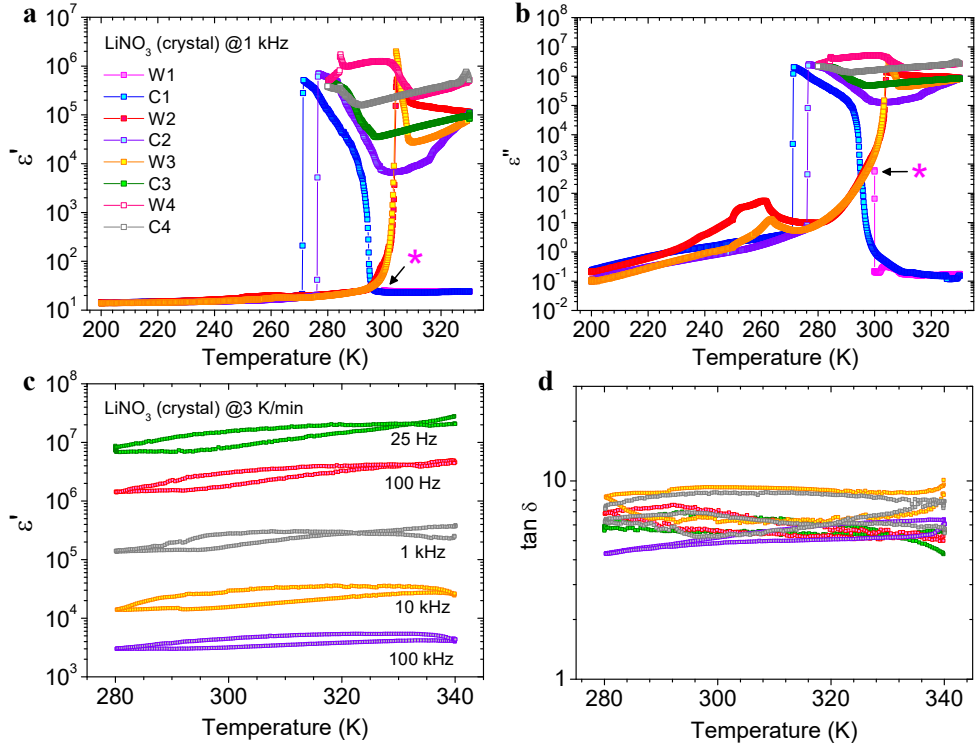

**Fig. 3.** (a) ε'($T$) and (b) ε"($T$) of single crystalline LiNO$_3$ under 1 kHz subjected to four warming and cooling cycles at 1 K/min. * symbols indicate the start of the measurements. (c) ε'($T$) and (d) tan δ of the same crystal under different frequencies.

### 3.2. Dielectric dispersion

Figures 4a to 4d show the complex dielectric constant as functions of frequency $f$ of LiNO$_3$ (single crystal) at 280 K, 310 K and 330 K in the high-temperature phase. Both ε'(ω) and ε"(ω) (where ω = 2π$f$) behaved monotonically between 20 Hz and 300 kHz, the limits of our impedance analyzer. The largest observed value of ε' was >2.6×10$^7$ at 20 Hz at 310 K. The insets in Figs. 4b and 4d show tan δ as a function of frequency. At frequencies below 1 kHz, tan δ was in the low single digits. Such a colossal value of ε' coupled to a relatively small tan δ while exhibiting thermal stability (from 280 K to 340 K) makes LiNO$_3$ an attractive candidate for insulation layers in nanoscale electronics.

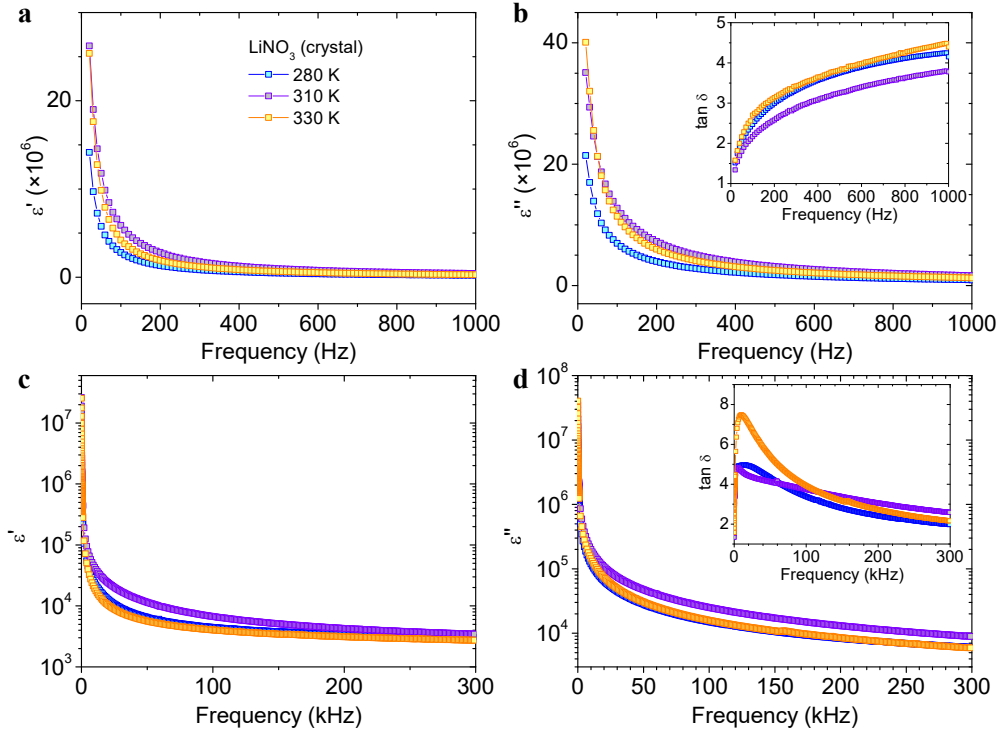

**Fig. 4.** Complex dielectric constant as a function of frequency of LiNO$_3$ at different temperatures in the (a), (b) near-static range and (c), (d) up to 300 kHz. Insets also show tan δ against frequency.

### 3.3. Electric polarization

Figure 5a shows the measured pyroelectric current of a single crystalline sample along the direction perpendicular to the *c*-axis. A *dc* electric field of $E = +555.6$ V/cm was applied to the sample at 330 K which was then cooled down to 260 K. The electric field was removed and the sample was warmed back to 330 K at the rate of 2 K/min while recording any measured current. This pyroelectric current $I(T)$ is shown in red in Fig. 5a. Immediately afterwards, $E = -555.6$ V/cm was applied to the sample and the abovementioned process was repeated with the recorded $I(T)$ shown in blue.

In both of the $I(T)$ curves a peak was observed near 305.5 K which was in good agreement with the ε'(*T*) and ε"(*T*) data (Figs. 2a and 3a). The area below each $I(T)$ curve is proportional to the electric polarization of the sample (shown in Fig. 5b). An unprecedented value of ~1,200 μC/cm$^2$ was recorded. As a comparison, the highest recorded polarizations have only been just above 100 μC/cm$^2$ [21-23].

The fact that the electric polarization can be switched according to an external applied electric field characterizes the low temperature ordered phase of LiNO$_3$ as ferroelectric. It should be noted that *P-E* measurements were also carried out at temperatures below $T_C$ but a remnant polarization was not observed. This is consistent with ferroelectrics of type-II as *P-E* hysteresis loops are usually only observable in type-I (displacive) ferroelectrics [24].

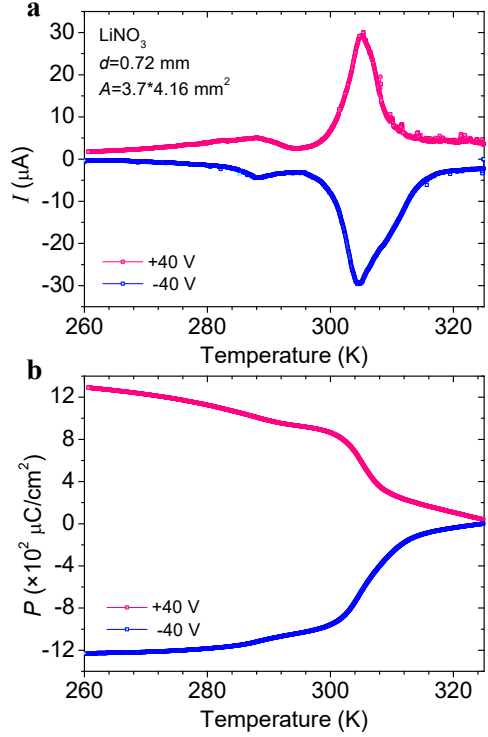

**Fig. 5.** (a) Measured pyroelectric current *I(T)* of LiNO$_3$ with thickness *d* and area *A* at 2 K/min. A *dc* electric field of ±555.6 V/cm was applied when the sample was cooled to 260 K before each measurement. (b) Derived electric polarization *P* from integration of *I(T)*.

### 3.4. Possible mechanism

As mentioned above, the magnitude of ε'(*T*) is proportional to the rate of libration of the NO$_3^-$. Hence, a natural question to ask is why do the NO$_3^-$ exhibit the type of observed dynamics? We first note that all NO$_3^-$ reside in an equilateral triangular lattice along the *ab*-planes (Fig. 6a). Moreover, the stacking of the *ab*-planes is in such a way that the positions of each NO$_3^-$ is exactly above the center of three NO$_3^-$ below. This means that each NO$_3^-$ is equidistant by $a_N$ to six NO$_3^-$ within the same *ab*-plane and equidistant by $c_N$ to six other NO$_3^-$ at different *z* (three above and three below). Figure 6b also lists the values of $a_N$ and $c_N$ of LiNO$_3$ at 380 K, 240 K and 100 K. At high temperatures (~380 K), the NO$_3^-$ librate independently from each other, however, at low temperatures (at low enough thermal fluctuations) neighboring anions become correlated because of conservation of angular momentum. For instance, if there were only two anions in the ground state with minimal fluctuations, as one NO$_3^-$ rotates by θ, then the other must do so by –θ (similar to Huygen's clocks) [25,26]. A problem appears as LiNO$_3$ is cooled because of its near-perfect triangular lattice. Once two adjacent NO$_3^-$ have become geared, a third equidistant NO$_3^-$ does not know whether to become correlated to the first or second NO$_3^-$ (Fig. 6b). Such geometric frustration is similar to that encountered in conventional magnetic systems where the spin of an unpaired electron equidistant to two others does not know whether to point up or down [27-29]. To remedy this situation, the lighter ion becomes off-centered because the high temperature phase is spatially disordered or because there are two types of anions such as in NH$_4$(SO$_4$)$_2$ [30]. However, Jahn-Teller-like distortions (at the molecular level) seem to be inhibited in LiNO$_3$ because of its unique near-perfect crystal lattice even at high temperatures. As neighboring NO$_3^-$ librate closer and closer to each other's natural frequency with decreasing temperature the system enters into a highly critical state since the NO$_3^-$ librate at each other's resonant frequency [31]. One solution to the sustained geometric frustration of the librations is that the NO$_3^-$ within entire *ab*-planes rotate in-phase while that of adjacent planes are offset by 180° (antiphase). This may be the reason behind the observed 40 K hysteresis between $T_W$ and $T_C$ as a lot of negotiating between neighboring NO$_3^-$ must take place in order for the entire system to become geared at the macroscopic level. The large hysteretic region is likely due to the system attempting to resolve its three-dimensional geometric frustration up until $T_C$ where the sharp drop-offs in $\varepsilon'(T)$ and $\varepsilon''(T)$ signal the resolving of the issue.

As the system is warmed up, the NO$_3^-$ librations remain geared up until $T_W$ where the vibrations of the Li$^+$ and/or NO$_3^-$ ions (thermal fluctuations) apparently begin to disrupt the macroscopic coherent state of the anions. The phase transition at $T_W$ is first order so the system first absorbs latent heat. This latent heat is usually used to change the structure of the system, however, the order-disorder phase transition in LiNO$_3$ is isostructural so the release of the absorbed latent heat seems to mostly be transferred onto the NO$_3^-$ libration modes which causes the huge spikes observed in $\varepsilon'(T)$ and $\varepsilon''(T)$ at $T_W$.

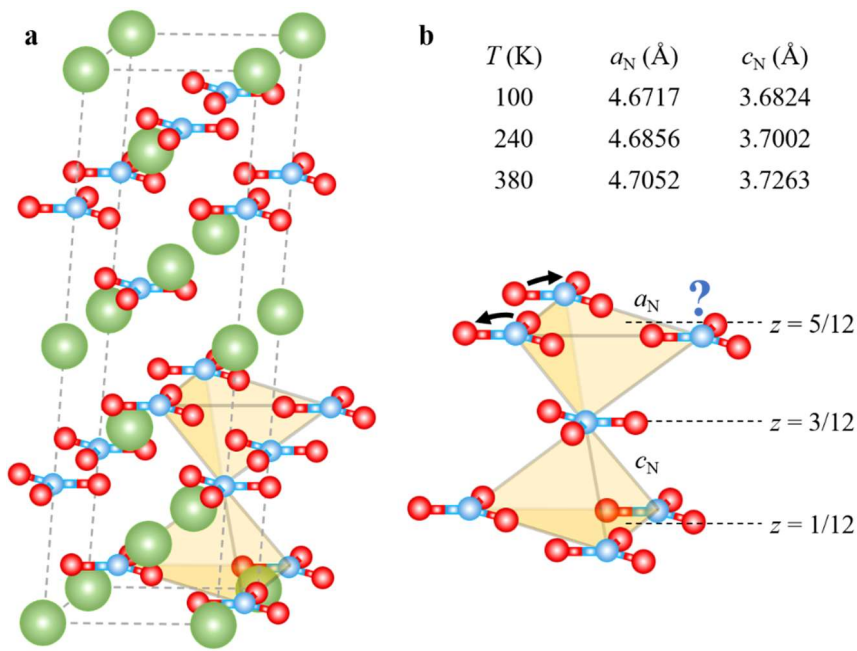

**Fig. 6.** (a) Crystal structure of LiNO$_3$. Parameters are available in the Supplementary Material. (b) NO$_3^-$ ions illustrating the inherent geometric frustration along the *ab*-planes and *c*-axis in the studied temperatures of 100, 240 and 380 K.

Regarding the observed ferroelectricity, the NO$_3^-$ anions are trigonal planar [32] and the Li$^+$ cations remain symmetrically distanced from the NO$_3^-$ in the ordered phase so a spontaneous polarization cannot be generated within the unit cell as all dipole moments cancel each other out. Alternatively, magnetic moments ordered into a spiral configuration can give rise to a spontaneous polarization [33-35]. This is possible because inversion of all magnetic moments along the propagation vector breaks the time-reversal and spatial-inversion symmetries. As each NO$_3^-$ rotates from $+\theta$ to $-\theta$, an equivalent magnetic moment perpendicular to the rotation plane is generated because the oxygen ions trace out current loops while the nitrogen atoms are near-stationary and the orbitals of the paired electrons cancel each other out. When the NO$_3^-$ ion rotates back to $+\theta$ from $-\theta$, the magnetic moment points along the opposite direction. If a phase shift $\xi$ exists between adjacent gearing *ab*-planes, then the magnetic moments of the NO$_3^-$ along the *c*-axis form spiral configurations. If the *c*-axis lattice constant is not a multiple of ξ, *i.e.* $c\,/\,\xi$ is incommensurate, then the ordered phase of LiNO$_3$ should exhibit a macroscopic polarization.

## 4. Conclusions

To conclude, while LiNO$_3$ possesses a colossal electric polarization (near 1,200 μC/cm$^2$), its applications for memory storage are limited because the temperature must be varied widely. On the other hand, a colossal dielectric response (~10$^6$) and large operating temperature range (280 to 340 K) enables LiNO$_3$ to potentially increase the performance of existing Class 2 ceramic capacitors, especially when the material is commercially available, inexpensive, light-weighted and environmentally friendly. The Supplementary Material file also shows ε' to be nearly independent of applied electric field from 0.6 to 25 V/cm. Lastly, a dissipation factor of $10^0$–$10^1$ at low frequencies also makes LiNO$_3$ attractive for use in sub-micron electronics.